# Energy-Aware Scheduling using Dynamic Voltage-Frequency Scaling

Masnida Emami [1,2], Yashar Ghiasi[1], Nasrin Jaberi [3]

[1] School of Science, Tarbiat Moaleem University, Karaj, Iran
[2] University College of Engineering and Technology Malaysia, Kuantan
[3] Payame-noor University, Najafabad

Nasrinjaberi60@yahoo.com

*Abstract*— The energy consumption issue in distributed computing systems has become quite critical due to environmental concerns. In response to this, many energy-aware scheduling algorithms have been developed primarily by using the dynamic voltage-frequency scaling (DVFS) capability incorporated in recent commodity processors. The majority of these algorithms involve two passes: schedule generation and slack reclamation. The latter is typically achieved by lowering processor frequency for tasks with slacks. In this article, we study the latest papers in this area and develop them. This study has been evaluated based on results obtained from experiments with 1,500 randomly generated task graphs.

## I. INTRODUCTION

To reduce energy consumption, various issues such as resource management in both software and hardware must be addressed. One issue in hardware resource management, that has a direct dependency on the number of transistors, is how to reduce power in processors. Energy consumption and task scheduling in multiprocessor systems can be addressed in two ways: (1) simultaneously, and (2) independently. In the simultaneous mode, tasks are (re)scheduled in a global cost function including energy saving and makespan to satisfy both energy and time constraints at the same time [1]. In fact because these constraints have opposite relation with each other, the result of such algorithms is typically a trade off between saving energy and less computation time depending on the weighting of these constraints in the global cost function. In the independent mode, a slack reclamation algorithm is used as a second pass to minimize the energy consumption of tasks in a schedule generated by a separate scheduler[2]. In this mode, energy and time constraints are independent and the existing scheduling algorithms in the literature are modified to become energy efficient. In the recent past, a lot of efforts have been put into the development of efficient power management mechanisms. The majority of them are hardware approaches, particularly for processors. Dynamic voltage-frequency scaling (DVFS) is perhaps the most appealing method incorporated into many recent processors. Energy savings with this method is typically achieved in the way that tasks are scheduled along processors slack times; this is based on the fact that the power consumption in CMOS circuits has direct relation with frequency and the square of voltage supply. In this case, the execution time and power consumption are controllable by switching between processor's frequencies and voltages. In [3], a linear combination of maximum and minimum frequencies in a processor is used for slack reclamation. This study then was completed in [4] where a combination of all frequencies are used. In this article we will study this paper and compare with the work in [5]. Moreover, the study in [6] indicates that if the relation between voltage and frequency is proportional, two adjunct frequencies are involved in optimal combination of frequencies; this observation is generalized in [7]. In this study, the problem of reducing energy consumption in the independent scheduling mode on multiprocessor systems is addressed –based on the paper in [4]. Unlike slack reclamation in most existing algorithms, this algorithm uses a linear combination of the processor frequencies to decrease energy consumption. For comparison, we refereed to the presented energy reduction algorithm in [3] and named it as Reference DVFS algorithm or RDVFS.

## II. THE ALGORITHM

The power consumption in CMOS circuits consists of two parts: (1) dynamic part that is mainly related to circuit switching energy, and (2) static part that addresses the circuit leakage power[8]. These two parts are formulated as:

$$P = \underbrace{CV^2 f}_{\substack{switching \\ energy}} + \underbrace{Kf}_{\substack{leakage \\ energy}}$$

Here $C$ is the effective switching capacitance, $V$ is the supply voltage, and $f$ is the clock speed[9]. When all frequencies are used, the problem of decreasing task energy, compared with the RDVFS algorithm (the reference DVFS is formulated as an optimization problem:

$$\begin{cases} E = \sum_{i=1}^{N}(CV_i^2 f_i + Kf_i)t_i \\ s.t. \\ 1. \sum_{i=1}^{N} t_i = T \\ 2. \sum_{i=1}^{N} f_i t_i = K \\ 3. t_i \geq 0, i = 1,2,\ldots,N \end{cases}$$

Where N in the number of processor frequencies, T is the maximum time a task can be stretch and K is the number of clock ticks required for executing the task. This optimization is applied for each task and tries to find the best frequencies. The same as [4], we name this algorithm MFS-DVFS.

## III. EVALUATION

The performance of MFS-DVFS was evaluated with randomly generated applications. For each application, a large number of variations in the number of tasks, number of processors and different characteristics of processors were applied to simulations. The processor is used in the simulations is Transmeta Crusoe TM-5800 CPU, extracted from [5]. A total of 1,500 experiments was performed with different task graphs on the five different numbers of processors in. The processing and release times of these randomly generated tasks varied between 5-10 and 2-20 time units from a uniform distribution, respectively. For each real-application graph, the same number of task graphs (ranging from 100 to 500 tasks) with three schedulers and on five sets of processors were investigated. Figure 1 shows the results of the proposed algorithm compare to the algorithm in [5] and the original scheduling.

## IV. CONCLUSION

Since most traditional static task scheduling algorithms in HPCS do not consider power management, the study in [4] addresses the energy issue with task scheduling and presents the MFS-DVFS algorithm. This algorithm adopted the DVFS technique, a recent advance in processor design, to reduce energy consumption. This work also studies how to use a linear combination of frequencies to reduce energy consumption on processors. Simulation in this article is based on results of 1500 randomly generated task graphs showing the effectiveness of the algorithm compared with the RDVFS algorithm (the reference DVFS algorithm).

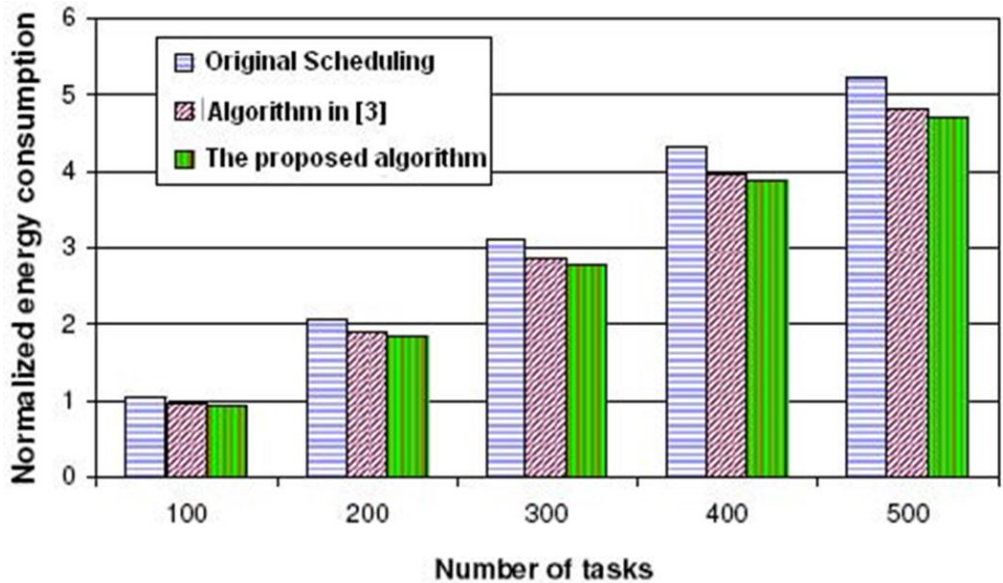

**Figure 1.** The normalized energy consumption on the number of tasks for the algorithm compared with [] and the original scheduling for three list schedulers: (a) The typical list scheduler (b) The list scheduler with Longest Processing Time first (LPT) and (c) The list scheduler with Shortest Processing Time first (SPT). The tasks in this figure come from 1500 random experiments averaged on 2, 4, 8, 16 and 32 processors.